\documentclass[floatfix,prb,aps,twocolumn,showpacs,preprintnumbers,amsmath,amssymb]{revtex4-1}
\usepackage{float}
\usepackage{amsfonts}
\usepackage{amsmath}
\usepackage{amssymb}
\usepackage{graphicx}
\usepackage{listings}
\usepackage{bm}
\usepackage{bbold}
\usepackage{braket}
\usepackage{todonotes}
\usepackage{hyperref}
\usepackage{url}
\usepackage[utf8]{inputenc}

\newcommand{\unit}[1]{\ensuremath{\,\mathrm{#1}}}
\newcommand{\R}{\ensuremath{\mathcal{R}}}
\renewcommand{\L}{\mathcal{L}}
\newcommand{\K}{\ensuremath{\mathcal{K}}}

\newcommand{\kun}{\ensuremath{\vec{k}_1}}
\newcommand{\kde}{\ensuremath{\vec{k}_2}}
\newcommand{\edem}{\ensuremath{\hat{\epsilon}_{p_2}^{-}}}
\newcommand{\eunp}{\ensuremath{\hat{\epsilon}_{p_1}^{+}}}

\begin{document}

\title{Pairwise summation approximation for Casimir potentials and its limitations}

\author{Anne-Florence Bitbol}
\altaffiliation{Present address: Laboratoire Mati\`{e}re et Syst\`{e}mes Complexes (MSC), CNRS UMR 7057, Universit\'{e} Paris Diderot}%
\affiliation{Laboratoire Kastler Brossel, ENS, UPMC, CNRS, F-75252 Paris, France}
\author{Antoine Canaguier-Durand}
\altaffiliation{Present address: ISIS - Universit\'{e} de Strasbourg.}
\affiliation{Laboratoire Kastler Brossel, ENS, UPMC, CNRS, F-75252 Paris, France}
\author{Astrid Lambrecht}
\affiliation{Laboratoire Kastler Brossel, ENS, UPMC, CNRS, F-75252 Paris, France}
\author{Serge Reynaud}
\affiliation{Laboratoire Kastler Brossel, ENS, UPMC, CNRS, F-75252 Paris, France}
\date{\today}

\begin{abstract}
We investigate the error made by the pairwise summation (PWS)
approximation in three geometries where the exact formula for the
Casimir interaction is known: atom-slab, slab-slab and sphere-slab
configurations. For each case the interactions are calculated
analytically by summing the van der Waals interactions between the
two objects. We show that the PWS result is incorrect even for an
infinitely thin slab in the atom-slab configuration, because of
local field effects, unless the material is infinitely dilute. In
the experimentally relevant case of dielectric materials, in all
considered geometries the error made by the PWS approximation is
much higher than the well-known value obtained for perfect
reflectors in the long-range regime. This error is maximized for
permittivities close to the one of Silicon.
\end{abstract}

\pacs{34.35.+a, 12.20Ds, 03.70.+k} \maketitle

\maketitle

\section{Introduction}

The Casimir effect is the universal attraction between two perfectly
reflecting plates in quantum vacuum due the zero-point fluctuations
of the electromagnetic field as first described in the seminal paper
by Casimir \cite{Casimir1948}. A quantum mechanical treatment of van
der Waals forces including the finite speed of light had already
been established by Casimir and Polder shortly before by means of
quantum electrodynamics \cite{CasimirPolderPR1948}. It is therefore
natural to wonder whether Casimir interactions between macroscopic
objects can be deduced from the van der Waals interactions between
their constituting atoms. A simple way of making this link would be
to sum the van der Waals interactions between all pairs of
constituting atoms, an approximation method known as \emph{pairwise
summation} (PWS). It is worth noting that attempts at explaining
macroscopic interactions between colloids by pairwise summing van
der Waals interactions were already carried out by de Boer and
Hamaker \cite{HamakerPh1937} ten years before the work of Casimir
and Polder.

However, the attractive PWS idea, that would reduce the Casimir
effect -- a physical effect of boundary conditions on the
electromagnetic vacuum field -- to a macroscopic resultant of
two-body interatomic forces, has not been fundamentally successful.
Indeed, PWS fails by construction to take into account any many-body
effect. Such effects, for instance the screening of electromagnetic
fields, are crucial in condensed matter physics, and have to be
accounted for when computing the Casimir effect between macroscopic
objects. Another fundamental cause of the insufficiency of PWS is
the existence of three-body van der Waals-type interactions
\cite{AxilrodJCP1943}. For permeable materials even the sign of the
interaction can be erroneous when calculated within PWS
\cite{KennethPRL2002}. Nevertheless, given the complexity of
performing exact calculations of the Casimir effect in different
geometries, the PWS approximation is still used nowadays since it
provides approximations to the true Casimir interactions
\cite{BlatterPRB1992,GirardPRB1994,DobsonPRB2008}. The results
obtained by PWS are then often empirically renormalized by a
corrective factor computed in a simpler case where the exact Casimir
interaction is known.

Limitations of PWS have been given in the past using a path integral
formulation for fluctuation-induced forces to study the
orientational dependence of PWS \cite{GolestanianPRE2000}, the force
between deformed plates \cite{EmigPRA2003} or within a perturbative
expansion in the dielectric contrast between arbitrarily shaped
bodies \cite{GolestanianPRL2005}. Recently first principle
calculations have obtained the PWS approximation as an asymptotic
approximation of the Casimir energy in the weakly-coupled\cite{MiltonPRL2008} or diluted limit\cite{LopezPRE2009}. In this paper, we
investigate the error made by PWS using the scattering approach
\cite{JaekelJP1991,LambrechtNJP2006,EmigPRL2007} in the atom-slab,
slab-slab, and sphere-slab geometries, three fundamental geometries
where the exact formula for the Casimir interaction is known. We
investigate which effects cause the failure of the PWS
approximation, and we study its error as a function of the material
and the slab thickness. In Sec.~\ref{sec:calc}, the interactions
between an atom and a slab, between two slabs and between a slab and
a sphere are calculated analytically by PWS at any distance between
the two interacting objects. Then, in Sec.~\ref{sec:lfe}, the
importance of local field effects as a cause of the failure of PWS
is emphasized. The influence of the material for bulk mirrors and
the effect of slab thickness in the long-range limit are studied in
Sec.~\ref{sec:material}. We summarize our findings in
Sec.~\ref{sec:ccl}.

\section{PWS calculation of Casimir interactions}
\label{sec:calc} PWS calculations are usually carried out in the
short-range and long-range limiting regimes
\cite{Milonni1994,MT1997,BezerraPRA2000,LopezPRE2009} where
Casimir-Polder interactions reduce to power laws. For the sake of
generality, we carry out the PWS calculation of the general
Casimir-Polder interaction in order to obtain PWS estimates of the
interactions between an atom and a slab, between two slabs, and between a sphere and a slab, at any
distance between them. In these three well-known geometries, the PWS
results can then be compared to the exact ones.

\subsection{Atom-slab geometry}
\label{ssec:as} We start from the van der Waals formula\cite{CasimirPolderPR1948} giving the
interaction energy between two atoms A and B of polarizabilities
$\alpha_A(\omega)$ and $\alpha_B(\omega)$, lying at a distance $d$:
\begin{align}
U_{a-a}(d)=-\frac{\hbar c}{\pi d^2} \int _0^{\infty} du \, u^4
\frac{\alpha_A (\imath cu) \alpha_B (\imath cu)}{(4\pi
\epsilon_0)^2} \, e^{-2ud} \nonumber \\ \times \left(
\frac{3}{u^4d^4}+ \frac{6}{u^3d^3} +\frac{5}{u^2d^2} +\frac{2}{ud}
+1\right) \label{atome_atome}
\end{align}
where the integral is written over imaginary frequencies,
$\omega=\imath cu$.

\begin{figure}[htb]
\includegraphics[width=0.25\textwidth]{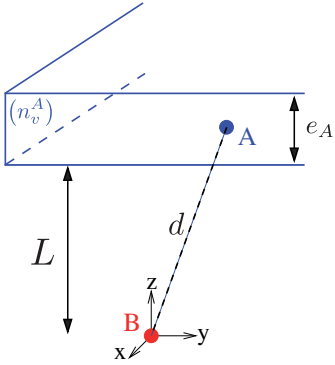}
\caption{Atom-slab geometry. \label{fig_slab_at}}
\end{figure}

We then consider a slab of thickness $e_A$ constituted of $n^A_v$
atoms per unit volume at a distance $L$ along the $z$ axis from an
isolated atom B (see Fig.~\ref{fig_slab_at}).

We compute the PWS interaction between the slab (s) and the atom (a)
by summing the van der Waals interactions (\ref{atome_atome})
between each atom A of the slab (s) and atom B:
\begin{align}
U^{\mathrm{PWS}}_{a-s}(L,e_A)=2\pi \,n^A_v \int_L^{L+e_A}
dz\int_0^\infty dr \,r\,U_{a-a}(d)
\end{align}
where $d=\sqrt{z^2+r^2}$. The integrations over $r$ and $z$ can be
carried out by parts after inverting them with the integration over
$u$, yielding the following result:

\begin{align}\label{atome_lame}
U^{\mathrm{PWS}}_{a-s}(L,e_A)&= \hbar c  n^A_v \int _0^{\infty} du \, u^3 \frac{\alpha_A (\imath cu) \alpha_B
(\imath cu)}{(4\pi \epsilon_0)^2} \nonumber \\
&  ~ ~ ~ ~ ~ ~  \times \left[ f(2uL)-f(2u(L+e_A))\right] \\
f(x)&=x\,\Gamma(0,x)-e^{-x}\left(1+\frac{4}{x^2}+\frac{4}{x^3}\right) \nonumber ~ ,
\end{align}
$\Gamma$ being the incomplete Gamma function $\Gamma :(a,x) \mapsto
\int^\infty_x dt \, t^{a-1}e^{-t}$.

A limiting case is the infinitely thick slab
($e_A\rightarrow\infty$), corresponding to a bulk material. We will
call this situation atom (a)- bulk plate (p) configuration in the
remainder of the paper. For this geometry we obtain the following
interaction potential within PWS:
\begin{align}
U^{\mathrm{PWS}}_{a-p}(L)= \hbar c \,n^A_v\int _0^{\infty} du \, u^3
\frac{\alpha_A (\imath cu) \alpha_B (\imath cu)}{(4\pi
\epsilon_0)^2} f(2uL) ~ . \label{atome_lamei}
\end{align}

\subsection{Slab-slab geometry}
\begin{figure}[htb]
\includegraphics[width=0.22\textwidth]{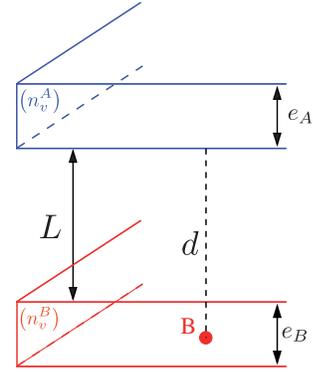}
\caption{Slab-slab geometry. \label{fig_slab_slab}}
\end{figure}

In order to compute the PWS estimate of the Casimir interaction
between two slabs, we simply sum the previous result
(\ref{atome_lame}) over atoms B constituting another slab with
thickness $e_{B}$ and number of atoms per unit volume $n_v^B$ in the
same way as before (see Fig.~\ref{fig_slab_slab}).

This procedure leads to the following PWS estimate of the Casimir
interaction between two slabs of thicknesses $e_A$ and $e_B$ at a
distance $L$, per unit surface:
\begin{multline}\label{lame_lame}
U^{\mathrm{PWS}}_{s-s}(L,e_A,e_B)=-\frac{\hbar c}{2}\,n^A_v n^B_v \int _0^{\infty} du \, u^2
\frac{\alpha_A (\imath cu) \alpha_B (\imath cu)}{(4\pi \epsilon_0)^2}  \\
 \times[\,g(2uL)+g(2u(L+e_A+e_B)) -g(2u(L+e_A))-g(2u(L+e_B))\,] 
\end{multline}
with 
\begin{align*}
 g(x)=\left(\frac{x^2}{2}-2\right)\Gamma(0,x) + e^{-x}\left(-\frac{x}{2}+\frac{1}{2}+\frac{2}{x}+\frac{2}{x^2} \right)
\end{align*}
a primitive function of $f(x)$ (see Appendix). When the slab
thicknesses become infinitely large we obtain for the interaction
between two bulk plates:
\begin{equation}\label{lamei_lamei}
U^{\mathrm{PWS}}_{p-p}(L)=-\frac{\hbar c}{2}\,n^A_v n^B_v \int
_0^{\infty} du \, u^2 \frac{\alpha_A (\imath cu) \alpha_B (\imath
cu)}{(4\pi \epsilon_0)^2} g(2uL) ~ .
\end{equation}

\subsection{Sphere-slab geometry}
\begin{figure}[htb]
\includegraphics[width=0.22\textwidth]{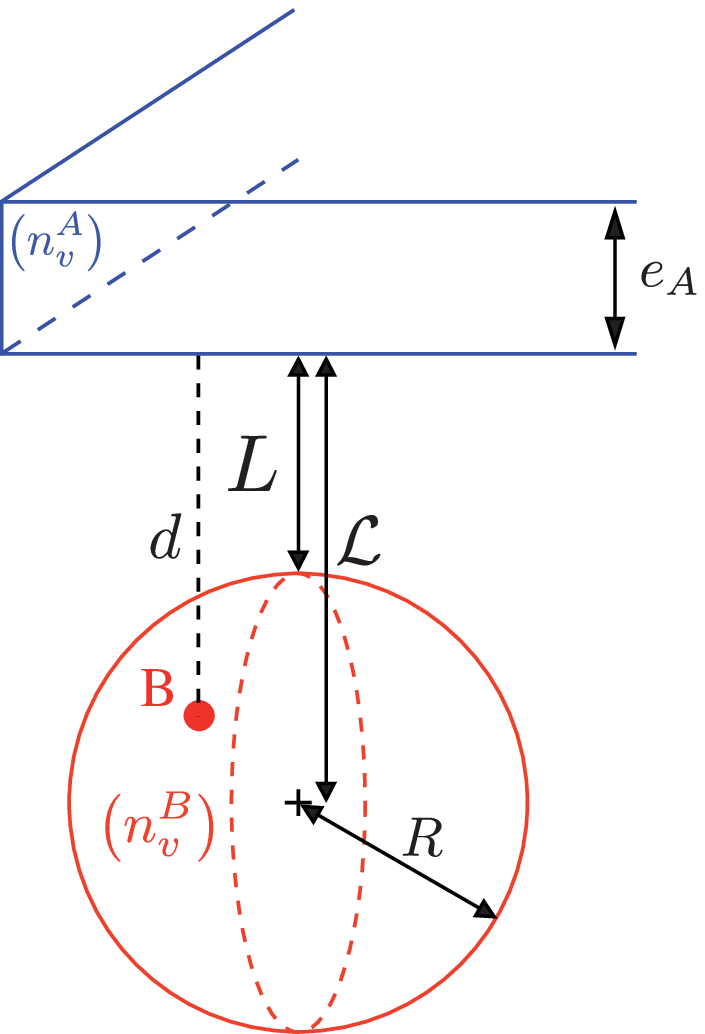}
\caption{Sphere-slab geometry. \label{fig_sphere_slab}}
\end{figure}

We next consider a third experimentally relevant situation where
object $A$ is (again) a slab of thickness $e_A$ and density $n_v^A$,
while object $B$ is a sphere of radius $R$ and density $n_v^B$. $L$
is the surface-surface distance of the two objects, while
$\mathcal{L}$ will be the center-to-plate distance
($\mathcal{L}=L+R$) (see Fig.~\ref{fig_sphere_slab}).

We compute the interaction between the two objects by summing the
atom-slab result given by Eq.(\ref{atome_lame}) for each atom of the
sphere $B$ which lies at a distance $d=\mathcal{L}+r$ from the slab,
with $r \in [-R,R]$:
\begin{align}
U^{PWS}_{sph-s} (\L,e_A) =  \int_{-R}^{R} dr~ \left( \pi (R^2-r^2)
n_v^B \right) U_{a-s}^{PWS} (\mathcal{L}+r,e_A) ~ .
\end{align}
After several integration by parts, it is possible to show that the
energy reads:
\begin{multline}
U^{PWS}_{sph-s} (\L,e_A)   = - \frac{\hbar c  \pi}{4} n_v^A n_v^B \int_0^\infty du~  \frac{\alpha_A(\imath cu) \alpha_B(\imath cu)}{(4\pi \epsilon_0)^2}  \\
\times \left\{ 2uR \left[ h\left( 2u(\mathcal{L} +e_A+R) \right) + h\left( 2u(\mathcal{L} +e_A-R) \right) \right. \right. \\
 - \left. h\left( 2u(\mathcal{L}+R) \right) - h\left( 2u(\mathcal{L} -R) \right) \right] \\
-  \left[ i\left( 2u(\mathcal{L} +e_A+R) \right) - i\left( 2u(\mathcal{L} +e_A-R) \right) \right. \\
\left. \left. - i\left( 2u(\mathcal{L}+R) \right) + i\left( 2u(\mathcal{L} -R) \right) \right] \right\}
\end{multline}
with 
\begin{align*}
h(x)  =  &\left( \frac{x^3}{6} -2t \right) \Gamma(0,x)+ e^{-x} \left(- \frac{x^2}{6} + \frac{x}{6} +\frac{5}{3} - \frac{2}{x} \right) \\
i(x)  =  &\left( \frac{x^4}{24} -x^2+2 \right) \Gamma(0,x) \\
& ~ ~ ~ ~ + e^{-x} \left(- \frac{x^3}{24} + \frac{x^2}{24} +\frac{11x}{12} - \frac{3}{4} \right)
\end{align*}
two successive primitives of
$g(x)$ (see Appendix). In the limiting case of
$e_A\rightarrow\infty$, we obtain:
\begin{multline}\label{sphere_lamei}
U^{PWS}_{sph-p} (\L)   =  \frac{\hbar c  \pi}{4} n_v^A n_v^B \int_0^\infty du~  \frac{\alpha_A(\imath cu) \alpha_B(\imath cu)}{(4\pi \epsilon_0)^2}  \\
\times \left\{ 2uR \left[   h\left( 2u(\mathcal{L}+R) \right) + h\left( 2u(\mathcal{L} -R) \right) \right] \right. \\
\left. - i\left( 2u(\mathcal{L}+R) \right) + i\left( 2u(\mathcal{L} -R) \right)  \right\} ~ .
\end{multline}

\section{Local field effects and PWS}
\label{sec:lfe} In Sec.~\ref{ssec:as}, we calculated the PWS
estimate of the atom-slab interaction by direct summation of the van
der Waals atom-atom interactions. We now show that the same result
can be derived within the scattering approach, which will make it
easier to compare analytically the PWS result and the exact Casimir
interaction in the atom-slab geometry.

\subsection{Equivalence between PWS and the summation of reflection matrices}
\label{ssec:eq} The Casimir interaction at zero temperature between
any two objects A and B at a distance $R$ along $z$ from one another
can be expressed as follows within the scattering approach to
Casimir forces \cite{JaekelJP1991,LambrechtNJP2006,EmigPRL2007}:
\begin{equation}
 U(R)= -\frac{\hbar c}{2 \pi} \int_{0}^{\infty} du \, \, \textrm{Tr} \, \ln \left(\mathbf{1}-\R_B \,  e^{-\K R}\, \R_A\,  e^{-\K R} \right) ~ .
\label{formule_g}
\end{equation}
In this formula, $\R_A$ and $\R_B$ are the reflection matrices of
the two interacting objects, while the two $e^{-\K R}$ factors are
propagation matrices. The term $\R_B \,  e^{-\K R}\, \R_A\,  e^{-\K
R}$ therefore corresponds to a complete roundtrip of the vacuum
field between the two objects A and B: the field propagates from B
to A before being reflected by A, then it propagates back to B and
is reflected by B.  The integration is on imaginary frequencies
$\omega=\imath cu$ of the electromagnetic field, while the trace Tr
runs over all independent field modes at a given frequency
$\omega=\imath cu$. Therefore, this formula takes into account the
contribution of every electromagnetic mode to the Casimir effect.

If one of the objects is an atom, only a small fraction of the field
is scattered back to object B and we may safely neglect multiple
reflections. Thus, if A or B -- or both -- are atoms, the general
formula (\ref{formule_g}) can be used at first order in $\R_B \,
e^{-\K R}\, \R_A\,  e^{-\K R}$, reducing to:
\begin{equation}
 U(R)= -\frac{\hbar c}{2 \pi} \int_{0}^{\infty} du \, \, \textrm{Tr} \, \left(\R_B \,  e^{-\K R}\, \R_A\,  e^{-\K R} \right) ~ .
\label{formule_g_ordre1}
\end{equation}
In particular, inserting the expressions of atom reflection matrices
(\ref{matrice_atome}) into the general Casimir interaction formula
at first order (\ref{formule_g_ordre1}) gives the van der Waals
interaction between two atoms (\ref{atome_atome}).

Let us now consider, as in Sec.~\ref{ssec:as}, the case of an
isolated atom B in front of a slab constituted of atoms A (see
Fig.~\ref{fig_slab_at}). Since formula (\ref{formule_g_ordre1}) is
linear in $\R_A$, it is equivalent to carry out a pairwise summation
of atom-atom interactions as in Sec.~\ref{ssec:as} or to sum the
reflection matrices of the constituting atoms in order to obtain the
reflection matrix of the slab. More precisely, if we write the slab
reflection matrix as $\R_s=\sum_{A}\R_A(\vec{R_A})$, where
$\vec{R_A}$ is the position of each atom A of the slab (the isolated
atom B being at the origin of coordinates), the atom-slab
interaction can be written using (\ref{formule_g_ordre1}):
\begin{eqnarray}
U_{a-s}(L)&=&-\frac{\hbar c}{2 \pi} \int_{0}^{\infty} du \, \textrm{Tr} \left(\R_B \,  e^{-\K L} \sum_{A}\R_A(\vec{R_A}) \, e^{-\K L} \right) \nonumber \\
&=& -\frac{\hbar c}{2 \pi} \sum_{A}\int_{0}^{\infty} du \, \textrm{Tr}  \left(\R_B \,  e^{-\K L}\, \R_A(\vec{R_A})\,  e^{-\K L} \right)\nonumber \\
&=&\sum_{A} U_{a-a}\left(R_A\right)  ~ .
\end{eqnarray}
We thus obtain the predicted result. It can also be understood
intuitively: summing the reflection matrices of the constituting
atoms amounts to considering each atom of the slab as an independent
scatterer. Each of these atoms then interacts independently with the
isolated atom through the electromagnetic field in an equivalent
manner to PWS.

\subsection{Calculation of slab reflection coefficients by summation of reflection matrices}
Having established that PWS is equivalent to summing the atom
reflection matrices in the atom-slab geometry, we will now compare
the exact reflection coefficients of a slab to the reflection
coefficients obtained by summing the reflection matrices of the
constituting atoms. For this, let us first calculate explicitly the
latter coefficients.

First, we write the position vector of an atom A of the slab as
$\vec{R}_A=(\vec{r_A},z_A=L+\zeta_A)$. Thus $\zeta_A$ ranges from 0
to $e_A$ (see Fig.~\ref{fig_slab_at}), while $\vec{r_A} \in
\mathbb{R}^2$. In order to compute
$\R_s=\sum_{A}\R_A(\vec{r_A},L+\zeta_A)$, we calculate a matrix
element of $\R_s$ in the plane wave basis. At a given frequency,
each plane wave is characterized by its two-dimensional transverse
(i.e. orthogonal to $z$) wave vector $\vec{k}$, its longitudinal
direction and its polarization $p$, transverse electric (TE) or
magnetic (TM). This description is complete since the modulus of the
longitudinal component $k^z$ of the wave vector can be deduced from
the relation $\omega=c\sqrt{|\vec{k}|^2+|k^z|^2}$.

Let us choose two plane waves $(\kun, p_1)$ and $(\kde, p_2)$, the
first one with longitudinal component in the + direction along $z$
and the second one with longitudinal component in the $-$ direction
along $z$. Their respective unit polarization vectors are therefore
noted $\eunp(\kun)$ and $\edem(\kde)$. The matrix element taken
between our two plane waves of the reflection matrix of an atom A at
$(\vec{0},L)$ is \cite{MessinaPRA2009}:
\begin{equation}
\langle \kde, p_2| \R_A |\kun, p_1 \rangle = -\frac{u^2 \alpha_A (\imath cu)}{2\kappa_2 \epsilon_0}\, \edem(\kde)\cdot \eunp(\kun) ~ .  \label{matrice_atome}
\end{equation}
In this formula, $\kappa_2$ is defined as: $|k_2^z|=\imath
\kappa_2$. We can write $\langle \kde, p_2|
\R_A(\vec{r_A},L+\zeta_A) |\kun, p_1 \rangle= \langle \kde, p_2|
\R_A |\kun, p_1 \rangle e^{-\imath (\kde-\kun)\cdot
\vec{r_A}}e^{-(\kappa_2-\kappa_1)\zeta_A}$, taking into account the
field's propagation between $(\vec{0},L)$ and each atom A situated
at $(\vec{r}_A, z_A=L+\zeta_A)$.

We are now able to compute the matrix element of
$\R_s=\sum_{A}\R_A(\vec{r_A},L+\zeta_A)$ between the two plane
waves:
\begin{eqnarray}
\langle \kde,p_2| \R_s |\kun, p_1 \rangle=\sum_{A}\langle \kde, &p_2|&\R_A(\vec{r_A},z_A)|\kun, p_1 \rangle \nonumber\\
= n_v\langle \kde, p_2| \R_A|\kun, p_1 \rangle \int_{\mathbb{R}^2}&d\vec{r_A}& e^{-\imath (\kde-\kun)\cdot \vec{r_A}} \nonumber\\
\int_0^{e_A} &d\zeta_A&e^{-(\kappa_2-\kappa_1)\zeta_A} ~ .
\end{eqnarray}
Once the integrals are calculated and $\langle \kde, p_2| \R_A|\kun,
p_1 \rangle$ is replaced by its explicit expression
(\ref{matrice_atome}), it yields:
\begin{eqnarray}
\langle \kde,p_2| \R_s |\kun, p_1 \rangle &=&n_v \, \frac{\pi u^2 \alpha_A (\imath cu)}{(4 \pi \epsilon_0) \,\kappa_1^2}\, \edem(\kun)\cdot \eunp(\kun)\nonumber\\
&&(2\pi)^2 \delta(\kde-\kun) \,\, (e^{-2\kappa_1e_A}-1) ~ .
\end{eqnarray}
The scalar products of polarization unit vectors can be easily
calculated by writing down the explicit form of these vectors as a
function of $\vec{k}$ and $k^z$ \cite{MessinaPRA2009}. Thus,
$\edem(\kun)\cdot \eunp(\kun)=0$ if $p_1\neq p_2$, while
$\hat{\epsilon}_{TE}^{-}(\kun)\cdot \hat{\epsilon}_{TE}^{+}(\kun)=1$
and $\hat{\epsilon}_{TM}^{-}(\kun)\cdot
\hat{\epsilon}_{TM}^{+}(\kun)=-(1+2\frac{k_1^2}{u^2})$. We finally
obtain:
\begin{eqnarray}
\langle &\kde&,p_2| \R_s |\kun, p_1 \rangle \nonumber\\
&=&(-1)^{\sigma(p_1)}\,(2\pi)^2 \delta(\kde-\kun) \, \delta_{p_1 p_2} \, r^{p_1}(\kun,u)
\label{matrice_R_plan}
\end{eqnarray}
where $\sigma(TE)=1$ and $\sigma(TM)=-1$, and where:
\begin{eqnarray}
r^{TE}_{\mathrm{slab}}(\kun,u)&=&n_v \, \frac{\pi u^2 \alpha_A (\imath cu)}{(4\pi\epsilon_0)\,\kappa_1^2} (e^{-2\kappa_1e_A}-1) \label{coef_te_som}\\
r^{TM}_{\mathrm{slab}}(\kun,u)&=&r^{TE}(\kun,u)\cdot
\left(1+2\frac{k_1^2}{u^2}\right) ~ . \label{coef_tm_som}
\end{eqnarray}

Equation (\ref{matrice_R_plan}) is totally general since it reflects
the invariance laws of specular reflection: the transverse wave
vector is conserved by specular reflection (this arises from the $x$
and $y$ invariance of the system through Noether's theorem), and TM
and TE polarization are eigenmodes of specular reflection.

On the contrary, the reflection coefficients (\ref{coef_te_som}) and
(\ref{coef_tm_som}) are specific to our summation approach. Using
these coefficients instead of the exact ones to compute the
atom-slab Casimir interaction is equivalent to using the PWS
approximation. We have proven this formally in Sec.~\ref{ssec:eq},
and it is straightforward to verify it explicitly by calculating the
atom-slab Casimir interaction using formula (\ref{formule_g_ordre1})
with the form of $\R_s$ we have just found: the result obtained is
exactly the one obtained by PWS, i.e. (\ref{atome_lame}).

\subsection{Comparison of the exact Casimir interaction and the PWS estimate in the atom-thin slab geometry}
In order to gain some insight into the error made by PWS, we now
compare the reflection coefficients calculated by summation in the
previous section to the exact ones. This is equivalent to comparing
the exact Casimir interaction and the PWS estimate in the atom-slab
geometry.

So as to obtain its exact reflection coefficients, the slab is
considered as a quantum optical network obtained by piling up a
vacuum/matter interface, propagation over a length $e_A$ in matter and
a matter/vacuum interface. Its transfer matrix is therefore the
product of the transfer matrices of the three elementary networks
constituting it. For a given polarization $p$=TE,TM the slab
reflection coefficients can be obtained from its transfer matrix
\cite{GenetPRA2003}.
\begin{equation} r^p_{\mathrm{slab}}=-\frac{\mathrm{sh}
\,\eta}{\mathrm{sh} (\eta+\theta^p)} ~ .  \label{coefs_exacts}
\end{equation}
In this formula, $\eta=e_A \kappa_m$ corresponds to propagation  in the
slab, with $\kappa_m=\sqrt{\epsilon u^2+k^2}$, while
$\theta^{TE}=\ln \big(\frac{\kappa_m+\kappa}{\kappa_m-\kappa}\big)$
and $\theta^{TM}=\ln
\big(\frac{\kappa_m+\epsilon\kappa}{\epsilon\kappa-\kappa_m}\big)$
are linked to the Fresnel vacuum/matter reflection coefficients
through the relation $r^p=-e^{-\theta^p}$. Here we have noted
$\imath \kappa=|k^z|$ the modulus of the longitudinal wave vector
and $k=|\vec{k}|$ the modulus of the transverse wave vector.

If the exact reflection coefficients (\ref{coefs_exacts}) are used
to calculate the atom-slab interaction from formula
(\ref{formule_g_ordre1}), the expression of the exact Casimir-Polder
interaction is obtained \cite{MessinaPRA2009}.

As both exact (\ref{coefs_exacts}) and PWS
(\ref{coef_te_som}-\ref{coef_tm_som}) reflection coefficients are
nonlinear function of the slab thickness $e_A$, we compare them in the
case of a thin slab ($e_A\rightarrow 0$). This limiting case is
interesting since one reason generally invoked to explain the errors
of PWS is that it fails to take into account the screening of the
electromagnetic field. If screening was the only problem of PWS,
this method would be most accurate for a thin slab.

A first order development in $e_A$ of the reflection coefficients obtained by summation (\ref{coef_te_som}-\ref{coef_tm_som}) gives:
\begin{eqnarray}
r^{TE}_{\mathrm{slab}}(\vec{k},u)&=&-2 \pi \,n_v \, e_A \,\frac{\alpha_A (\imath cu)}{4\pi\epsilon_0}\frac{u^2 }{\kappa} \label{r_te_s}\\
r^{TM}_{\mathrm{slab}}(\vec{k},u)&=&2 \pi \,n_v \, e_A
\,\frac{\alpha_A (\imath cu)}{4\pi\epsilon_0}\frac{u^2
}{\kappa}\left(1+2\frac{k^2}{u^2} \right) \label{r_tm_s} ~ .
\end{eqnarray}
Similarly, the exact reflection coefficients (\ref{coefs_exacts}) become, at first order in $e$:
\begin{eqnarray}
r^{TE}_{\mathrm{slab}}(\vec{k},u)&=&-\frac{e_A(\epsilon-1)u^2}{2\kappa} \label{r_te_e}\\
r^{TM}_{\mathrm{slab}}(\vec{k},u)&=&-\left( 1+\frac{\epsilon+1}{\epsilon}\frac{k^2}{u^2} \right) \frac{e_A(\epsilon-1)u^2}{2\kappa} ~ .
\label{r_tm_e}
\end{eqnarray}

We notice that, at first order in $e_A$, the two TE reflection
coefficients (\ref{r_te_s}) and (\ref{r_te_e}) are identical only if
$\epsilon=1+4 \pi n_v \, \frac{\alpha_A(\imath cu)}{4
\pi\epsilon_0}$, while the two TM reflection coefficients
(\ref{r_tm_s}) and (\ref{r_tm_e}) are identical only if
$\frac{\epsilon+1}{\epsilon}=2$, that is to say if $\epsilon-1$
vanishes. Thus, we can say that PWS gives exact results in the
atom-thin slab geometry only if $\epsilon=1+4\pi\,n_v \,
\frac{\alpha_A(\imath cu)}{4 \pi\epsilon_0}$ with $4\pi\,n_v \,
\frac{\alpha_A(\imath cu)}{4 \pi\epsilon_0}\ll1$, that is to say
only in the case when the material of the slab is infinitely
diluted.

This condition found in the limit of thin slabs is very general. In
fact it can proven to be sufficient for any slab thickness: the
condition $\epsilon(\imath c u)=1+4\pi\,n_v \, \frac{\alpha(\imath c
u)}{4 \pi\epsilon_0}$ with $4\pi\,n_v \, \frac{\alpha(\imath c u)}{4
\pi\epsilon_0}\ll1$ is the first-order approximation of the
Clausius-Mossotti relation:
\begin{align}\label{clausius}
\frac{\epsilon(\imath c u)-1}{\epsilon(\imath c u)+2}=\frac{4\pi}{3} n_v \frac{\alpha(\imath c u)}{4 \pi \epsilon_0}
\end{align}
 when $n_v\,\alpha(\imath c u)$ is very small, i.e. at the diluted limit. If we insert this first-order approximation to the Clausius-Mossotti relation in the exact
 reflection coefficients (\ref{coefs_exacts}), we obtain the reflection coefficients found by summation (\ref{coef_te_som}-\ref{coef_tm_som}) for any slab thickness.

Replacing the Clausius-Mossotti relation by its first-order
approximation amounts to neglecting local field effects. Therefore
our analysis indicates that local field effects are crucial in the
errors made by PWS. Moreover, the fact that the condition of
validity of PWS is not more restrictive for a thick slab than for a
very thin one suggests that screening is not the most important of
the effects that cause the errors of PWS.

\section{Influence of the material in the long-range limit}
\label{sec:material} The formal comparison of the Casimir
interaction and its PWS estimate in the atom-slab geometry has
enabled us to gain some insight into the reasons why PWS is not
exact, emphasizing the role of local field effects among the
different many-body effects that are not taken into account in PWS.
However, such formal comparisons are quite difficult to carry out in
general, and it is of great practical importance to know the
magnitude of the error made by approximations such as PWS.

The results that are generally cited about the errors of PWS involve
perfect reflectors \cite{Milonni1994,MT1997}. In the long-range
(retarded) limit, the ratio between the pairwise approximation and
the exact value of the atom-slab interaction is 1.15 if the slab is
a perfect reflector. In the slab-slab geometry, this ratio is 0.80
\cite{Milonni1994}. As the perfect reflector is an idealization, we
may wonder how this ratio changes for real materials. In this
section, we estimate numerically the ratio between the pairwise
approximation and the exact value of the Casimir interaction for
dielectric materials in the atom-slab and slab-slab geometries. In
order to achieve model-independent results we restrict ourselves to
the long-range limit.

\subsection{Atom-slab interaction}
Let us in a first step suppose that the slab thickness $e_A$ is much
larger than $L$. The influence of the slab thickness will be
addressed later-on.

Besides we suppose that the distance $L$ between the atom and the
slab is much larger than any relevant intrinsic characteristic
distance, such as the wavelengths corresponding to the atomic
transitions of the isolated atom, and any wavelength characteristic
of the permittivity of the slab material. Practically, for a typical
atom and a silicon slab with permittivity modeled by a Drude-Lorentz
formula \cite{PirozhenkoPRA2008}, this means that $L>1\unit{\mu m}$.

The exact atom-bulk plate interaction can be expressed thanks to
formula (\ref{formule_g_ordre1}), where $\R_P$ is the slab
reflection matrix (\ref{matrice_R_plan}) with the exact reflection
coefficients (\ref{coefs_exacts}). The latter reduce to the bulk
Fresnel reflection coefficients, since we are dealing with infinite
slabs. The result reads \cite{MessinaPRA2009}:
\begin{eqnarray}
U_{a-p}(L)&=& \frac{\hbar c}{2\pi} \int_{0}^{\infty} du \,  u^2 \frac{\alpha_B(\imath cu)}{4\pi\epsilon_0}\int_u^\infty d\kappa \, e^{-2\kappa L} \nonumber\\
&&\left( r^{TE}+\left(2\frac{\kappa^2}{u^2}-1\right)r^{TM}\right)
\label{atome_surf_g}
\end{eqnarray}
where  $r^{TE}=\frac{\kappa-\kappa_m}{\kappa+\kappa_m}$ and $r^{TM}=
\frac{\kappa_m-\epsilon\kappa}{\epsilon\kappa+\kappa_m}$, with
$\kappa_m=\sqrt{u^2\epsilon(\imath cu)+k^2}$. In the long-distance
limit, the propagation factor $e^{-2\kappa L}$ in the atom-plate
Casimir interaction formula (\ref{atome_surf_g}) vanishes except at
small frequencies $\omega=\imath cu$, so that the polarizability of
the isolated atom $\alpha_B(\imath cu)$ and the permittivity
$\epsilon(\imath cu)$  can be replaced by their static values
$\alpha_B(0)$ and $\epsilon(0)$.

In the PWS approximation, the interaction will thus be given by
formula (\ref{atome_lamei}). As before, the propagation factor
$e^{-2uL}$ allows for replacement of $\alpha_A(\imath cu)$ and
$\alpha_B(\imath cu)$ by $\alpha_A(0)$ and $\alpha_B(0)$ in this
formula in the long distance limit. Explicit integration then shows
that formula (\ref{atome_lamei}) reduces to:
\begin{equation}
U^{\mathrm{PWS}}_{a-p}(L)=-\frac{23}{40} \, \frac{\hbar c
\,n^A_v}{L^4} \, \frac{\alpha_A (0) \alpha_B (0)}{(4\pi
\epsilon_0)^2} ~ . \label{atome_lamei_ret}
\end{equation}
This long-distance limit of formula (\ref{atome_lamei}) can be found
by simple pairwise summation of the retarded Casimir-Polder
interaction \cite{Milonni1994}.

We notice that in the long-distance limit, $\alpha_B (0)$ is a
simple multiplicative factor in both the exact Casimir interaction
and its PWS estimate. Therefore, in this limit, the ratio
$U^{\mathrm{PWS}}_{a-p}/U_{a-p}$ does not depend on any property of
the isolated atom. Besides, in order to compare the PWS estimate to
the exact Casimir interaction, $\alpha_A(0)$ is expressed as a
function of $\epsilon(0)$ thanks to the Clausius-Mossotti relation
(\ref{clausius}).

\begin{figure}[htb]
\includegraphics[width=0.48\textwidth]{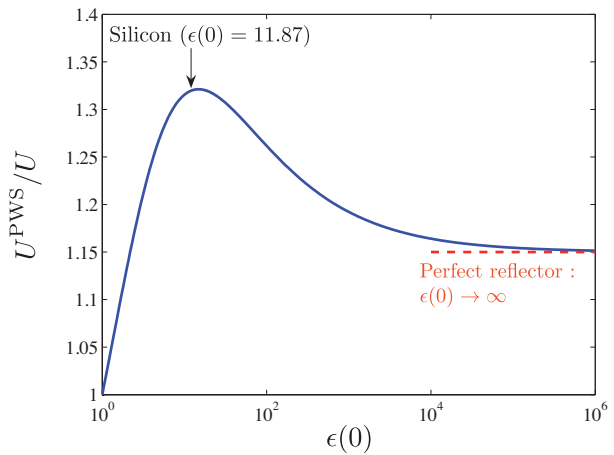}
\caption{Ratio of the PWS estimate of the atom-plate Casimir
interaction and its exact value in the long-range limit. This ratio
is traced as a function of the static
permittivity.\label{ratio_slab_at}}
\end{figure}

It is now possible to compute the ratio
$U^{\mathrm{PWS}}_{a-p}/U_{a-p}$ in the long range  limit as a
function of the only remaining material parameter $\epsilon(0)$. The
result obtained is traced in Fig.~\ref{ratio_slab_at}.

We observe on Fig.~\ref{ratio_slab_at} that the PWS method becomes
correct when $\epsilon(0)\rightarrow 1$, that is to say when the
bulk matter becomes infinitely diluted. This is consistent with the
conclusions of Sec.~\ref{sec:lfe}. Besides, we find the well-known
result $U^{\mathrm{PWS}}_{a-p}/U_{a-p}=\frac{23}{20} \simeq 1.15$
corresponding to the perfect reflector in the limit
$\epsilon(0)\rightarrow \infty$ \cite{Milonni1994,MT1997}. An
unexpected and interesting result is that the ratio
$U^{\mathrm{PWS}}_{a-p}/U_{a-p}$ does not evolve monotonically
between these two limits, but has a maximum at $\epsilon(0)\simeq
14.9$, where $U^{\mathrm{PWS}}_{a-p}/U_{a-p} \simeq 1.321$.
Moreover, this maximum corresponds to realistic dielectric
materials. For instance, we show on Fig.~\ref{ratio_slab_at} that
silicon is very close to it, with $\epsilon(0)=11.87$ and
$U^{\mathrm{PWS}}_{a-p}/U_{a-p} \simeq 1.319$. PWS thus makes a
larger error for dielectrics such as silicon than for a perfect
reflector if the atom-bulk plate interaction is considered. The
relative error can reach about 30\%.

We now consider a finite thickness $e_A$ for the slab and study its
effect on the accuracy of the PWS method.  We start from the derived
expression (\ref{atome_lame}) for the energy between the slab and
the atom, and then apply the long-distance limit so that the
polarizability of the isolated atoms  can be replaced by their
static values. It leads to a result similar to the bulk case
(\ref{atome_lamei_ret}), except for a thickness-dependent factor:
\begin{equation}
U^{\mathrm{PWS}}_{a-s}(L,e_A)=U^{\mathrm{PWS}}_{a-p}(L) \left( 1 - \frac{1}{\left(1+\frac{e_A}{L} \right)^4}\right) ~  . \label{atome_slab_ret}
\end{equation}

\begin{figure}[htb]
\includegraphics[width=0.48\textwidth]{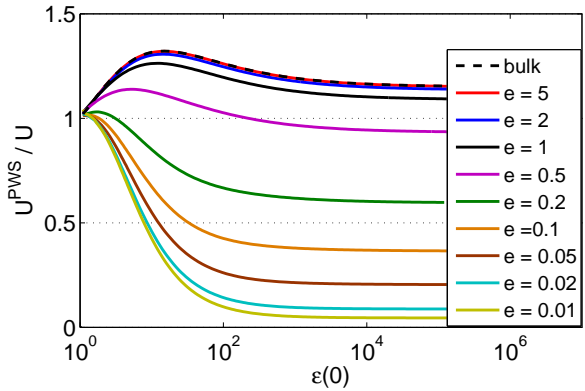}
\caption{Ratio of the PWS estimate of the atom-slab Casimir
interaction and its exact value in the long-range limit, for various
values of the relative thickness $e=e_A/L$. The case of a bulk plane
is recalled with a dashed-line.\label{ratio_slab_at_e}}
\label{ratio_slab_at_thickness}
\end{figure}

For the computation of the exact Casimir energy, we use
Eq.(\ref{atome_surf_g})  with the thickness-dependent Fresnel
coefficients introduced in Eq.(\ref{coefs_exacts}). The obtained
ratio is traced in Fig.~\ref{ratio_slab_at_e} for various values of
the relative thickness $e=e_A/L$ which is dimensionless.
We observe that the ratio of the PWS estimation over the exact
result is also strongly dependent  on the thickness of the slab
compared to the distance between the two objects: while for a large
thickness compared to the distance $(e\gg 1)$ the bulk plate case is
recovered for any static permittivity, the ratio decreases
dramatically when the thickness decreases. At the limit of infinite
permittivity, or perfectly reflecting objects, the ratio goes as
expected to the thickness-dependent ratio $\left( 1 - \left(1+e_A/L
\right)^{-4} \right)$.

\subsection{Two slabs}
Let us now carry out the same study for two bulk materials. In this
geometry, the whole formula (\ref{formule_g}) has to be used since
multiple reflections are not negligible. Using equation
(\ref{matrice_R_plan}) for the reflection matrix of each plate, this
formula gives the energy per unit area:
\begin{equation}
U_{p-p}(L)= \frac{\hbar c}{2 \pi}  \int_{0}^{\infty} du \, \,
\textrm{Tr} \, \ln \left(1-r_B \, r_A\,  e^{-2\kappa L} \right)
\label{formule_g_spec}
\end{equation}
This formula is only valid for specular reflection. The general
formula (\ref{formule_g}) for the Casimir interaction between any
two objects was in fact first obtained as a generalization of the
specific formula (\ref{formule_g_spec}) \cite{LambrechtNJP2006}.
After writing explicitly the trace over field modes, formula
(\ref{formule_g_spec}) becomes:
\begin{eqnarray}
U_{p-p}(L)= \frac{\hbar c}{4\pi^2} \int_{0}^{\infty} &du & \int_u^\infty d\kappa \, \kappa \,\big[\,\ln \left( 1-(r^{TE})^2 e^{-2\kappa L} \right)\nonumber\\
&+& \ln \left( 1-(r^{TM})^2 e^{-2\kappa L} \right)\,\big]
\label{slab_slab}
\end{eqnarray}
with the bulk reflection coefficients $r^{TE}$ and $r^{TM}$ such as
defined after eqn. (\ref{atome_surf_g}). We have assumed here that
both plates are identical. In the case of perfect reflectors
($r^2=1$), formula (\ref{slab_slab}) reduces after some calculations
to the original Casimir formula \cite{Casimir1948} $U_{p-p}(L)=-
\frac{\hbar c\pi^2}{720L^3}$.

In the PWS approximation, the interaction between two plates at any
distance has been computed in Sec.~\ref{sec:calc}, formula
(\ref{lamei_lamei}). In the long distance limit the polarizabilities
can be replaced by their static values, so that the general PWS
result (\ref{lamei_lamei}) can be simplified:
\begin{equation}
U^{\mathrm{PWS}}_{p-p}(L)=-\frac{23}{120} \, \frac{\hbar c \,n_v^A
n_v^B}{L^3} \, \frac{\alpha_A (0) \alpha_B (0)}{(4\pi \epsilon_0)^2}
\label{lamei_lamei_ret}
\end{equation}
This long-distance limit of formula (\ref{lamei_lamei}) can also be
found by simple pairwise summation of the retarded Casimir-Polder
interaction \cite{Milonni1994}. Here we have $\alpha_A=\alpha_B$ and
$n_v^A=n_v^B$ since both plates are supposed to be identical.

\begin{figure}[htb]
\includegraphics[width=0.48\textwidth]{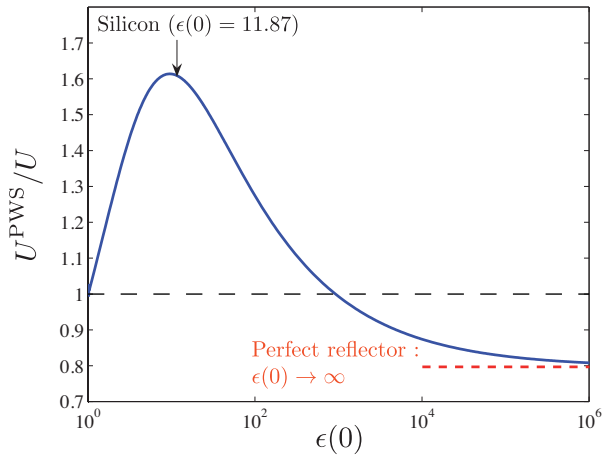}
\caption{Ratio of the PWS estimate of the plate-plate Casimir
interaction and its exact value in the long-range limit. This ratio
is traced as a function of the material's static
permittivity.\label{ratio_slab_slab}}
\end{figure}

In order to compare the exact Casimir interaction and its PWS
estimate, we use once more the Clausius-Mossotti relation
(\ref{clausius}) to express $\alpha(0)$ as a function of
$\epsilon(0)$. We now compute the ratio
$U^{\mathrm{PWS}}_{p-p}/U_{p-p}$ in the long range limit using the
same method as in the previous section. The result obtained is
traced as a function of $\epsilon(0)$ in Fig.~\ref{ratio_slab_slab}.

While the global aspect of the ratio
$U^{\mathrm{PWS}}_{p-p}/U_{p-p}$ in Fig.~\ref{ratio_slab_slab} is
similar to the one plotted in Fig. \ref{ratio_slab_at} (atom-bulk
plate interaction), the error of PWS in the two plates geometry is
found to be twice as large, reaching now up to 60\%.

Moreover, a striking feature in Fig.~\ref{ratio_slab_slab} is that
the ratio $U^{\mathrm{PWS}}_{p-p}/U_{p-p}$ is larger than one for
most dielectric materials ($\epsilon(0) \lesssim 100$), while it is
smaller than one for the perfect reflector
($U^{\mathrm{PWS}}_{p-p}/U_{p-p}=\frac{621}{8\pi^4} \simeq 0.797$).
Thus, the perfect reflector result does not provide a correct
explanation as to the reasons of failure of PWS in most materials.
This observation illustrates the complexity of the error made by
PWS: the PWS estimate can be either smaller or larger than the exact
Casimir interaction depending on the material for a given geometry.
Furthermore, screening of electromagnetic fields by slab atoms turns
out not be the main responsible for the error as we would then
expect the ratio $U^{\mathrm{PWS}}_{p-p}/U_{p-p}$ to be larger than
one for all values of $\epsilon(0)$, which is not the case here. As
we have argued in Sec.~\ref{sec:lfe}, local field effects seem to
play a crucial part in this error.

We now consider finite thicknesses $e_A, e_B$ for the slabs and
study the effect of these  parameters on the accuracy of the PWS
method. We start from the derived expression (\ref{lame_lame}) for
the energy between the slabs, and then apply the long-distance limit
so that the polarizability of the isolated atoms  can be replaced by
their static values. It leads to a result similar to the bulk case
(\ref{lamei_lamei_ret}), except for a thickness-dependent factor:
\begin{multline}\label{slab_slab_ret}
U^{\mathrm{PWS}}_{s-s}(L,e_A,e_B)=U^{\mathrm{PWS}}_{p-p}(L)    \\
\times \left( 1 - \frac{1}{\left(1+\frac{e_A}{L} \right)^3} - \frac{1}{\left(1+\frac{e_B}{L}  \right)^3} + \frac{1}{\left(1+\frac{e_A}{L}  + \frac{e_B}{L}  \right)^3} \right) ~  .
\end{multline}

\begin{figure}[htb]
\includegraphics[width=0.5\textwidth]{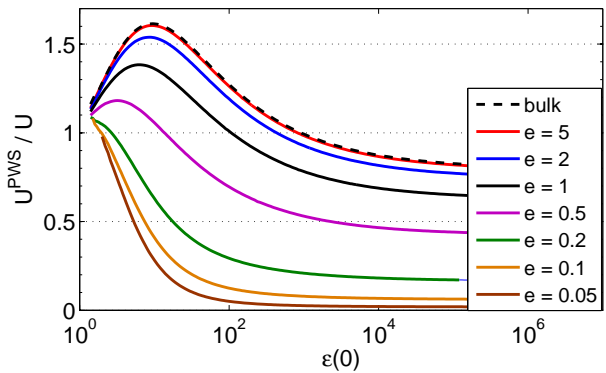}
\caption{Ratio of the PWS estimate of the slab-slab Casimir
interaction and its exact value in the long-range limit, for various
values of the relative  thickness $e=e_A/L=e_B/L$. The case of two
bulk planes is recalled with a
dashed-line.\label{ratio_slab_slab_e}}
\end{figure}

For the computation of the exact Casimir energy, we use
Eq.(\ref{slab_slab}) with the  thickness-dependent Fresnel
coefficients introduced in Eq.(\ref{coefs_exacts}). For simplicity
we will only consider the case where $e_A=e_B$. The obtained ratio
is traced in Fig.~\ref{ratio_slab_slab_e} for various values of the
relative thickness $e=e_A/L=e_B/L$, as in the atom-slab case in
Fig.~\ref{ratio_slab_at_thickness}.

We observe that the ratio of the PWS estimation over the exact
result is also strongly  dependent on the thickness of the slabs
compared to the distance between them: while for a large thickness
compared to the distance $(e\gg 1)$ the bulk plate case is recovered
for any static permittivity, the ratio decreases dramatically when
the thickness decreases. At the limit of infinite permittivity, or
perfectly reflecting objects, the ratio goes as expected to the
thickness-dependent ratio $\left( 1 - 2 \left(1+ e \right)^{-3} +
\left(1+ 2e \right)^{-3}\right)$.

\subsection{Sphere-bulk plate geometry}

We finish this study with the sphere-slab configuration, where a
sphere of radius $R$ has its center at a distance $\L$ from an
infinitely thick slab. In this geometry the exact quantity for the
Casimir energy $U_{sph-p}$ has to be derived from the general
scattering formula (\ref{formule_g}), with $\R_S$ and $\R_P$ the
reflection operators on the sphere and on the plane, respectively.
These operators can be expressed thanks to the bases of planar and
spherical waves, as discussed in details in
\cite{NetoPRA2008,CanaguierPRA2010}. In the present study, we
consider the long-range limit, in the sense that the distance
between the sphere and the infinite slab $L=\L-R$ is much larger
than any wavelength $\lambda$ characteristic of the permittivity
$\epsilon$ of the objects material. As in the previous cases the
permittivity and polarizability in the reflection operators $\R_S$
and $\R_P$ can then be replaced by their static value $\epsilon(0)$.

In the PWS approximation, the interaction between a sphere and a
plate has been computed in Sec.~\ref{sec:calc}, formula
(\ref{sphere_lamei}). In the long-range limit the general PWS result
(\ref{sphere_lamei}) can be simplified to:
\begin{equation}
U^{\mathrm{PWS}}_{sph-p}(L)=-\frac{23}{30} \, \frac{\hbar c \pi R^3
\,n_v^A n_v^B}{(\L^2-R^2)^2} \, \frac{\alpha_A (0) \alpha_B
(0)}{(4\pi \epsilon_0)^2} \label{lamei_lamei_ret}
\end{equation}
Here we have $\alpha_A=\alpha_B$ and $n_v^A=n_v^B$ since the materials for the sphere and the slab are supposed to be identical.

In order to compare the exact Casimir interaction and its PWS
estimate, we use once more the Clausius-Mossotti relation
(\ref{clausius}) to express $\alpha(0)$ as a function of
$\epsilon(0)$. We now compute the ratio
$U^{\mathrm{PWS}}_{sph-p}/U_{sph-p}$ in the long range limit using
the same method as in the previous sections. The result obtained is
traced as a function of $\epsilon(0)$ in Fig.~\ref{ratio_slab_slab},
for several values of the additional geometrical parameter
$\frac{L}{R}$.

\begin{figure}[htb]
\includegraphics[width=0.48\textwidth]{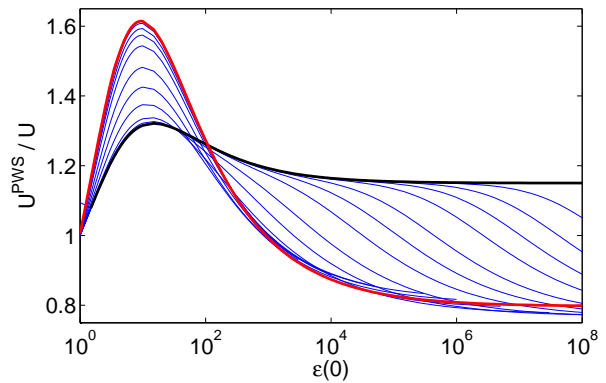}
\caption{Ratio of the PWS estimate of the sphere-slab Casimir interaction and its exact value in the case of infinitely thick slabs in the long-range limit, for various geometrical parameters $\frac{L}{R}$ from 0.02 to 5000 (blue curves). This ratio is traced as a function of the static permittivity of the slab material. The black and red curves are a recall of the atom-slab and slab-slab cases, previously presented in Figs. \ref{ratio_slab_at} and \ref{ratio_slab_slab}, respectively. \label{ratio_sphere_slab}}
\end{figure}

The first observation is that the ratio
$U^{\mathrm{PWS}}_{sph-p}/U_{sph-p}$ is dependent on the parameter
$\frac{L}{R}$: for small values (i.e. very large spheres) this ratio
tends to the plate-plate result, recalled by the red curve. On the
opposite, for large values of $\frac{L}{R}$ (i.e. small spheres) the
curves converge to the atom-plate result, presented with a dark
curve, except for very large values of $\epsilon$.  Indeed, the
small-sphere limit ($R \ll L$) for perfect mirrors yields a ratio
$\frac{23}{30}\simeq0.77$, while the $\epsilon(0)\rightarrow \infty$
limit for an atom in front of a slab leads to a ratio
$\frac{23}{20}=1.15$. This shows that the two limits $(R \ll L)$ and
$\left( \epsilon(0)\rightarrow \infty \right)$, of a small sphere and of a perfect reflector, do not commute. The ratio $\frac{3}{2}$ is typical for this non-commutativity\cite{CanaguierPRA2010} and can be traced back to the small parameter-limit for the Mie coefficient $b_\ell$, or in other terms, to the impossibility to have a magnetic point-like dipole.

\section{Conclusion}
\label{sec:ccl} We have investigated the error made by the pairwise
summation method in three geometries where the exact formula for the
Casimir interaction is known, the atom-slab, the slab-slab and the
sphere-slab geometry. The scattering approach to the Casimir effect
has enabled us to show, through an analytical comparison of
reflection coefficients, that the PWS result is incorrect even for
an infinitely thin slab interacting with an atom, unless the
material is infinitely dilute. This analysis has stressed the
fundamental importance of local field effects among the many-body
effects that PWS fails to take into account.

We have then studied the influence of the material on the error made
by PWS. This study has shown that the error made by the PWS method
is much higher in the experimentally relevant case of dielectric
materials such as silicon than for perfect reflectors. We have
reached this conclusion in the long range limit, both in the
atom-slab and in the slab-slab geometry. The existence of a maximum
in the error made by PWS for the permittivity of usual dielectrics
is not easy to understand intuitively and sheds light on the
complexity of the error made by PWS. So does the fact that PWS can
underestimate or overestimate the Casimir interaction in the
slab-slab geometry depending on the slab permittivity. Although the
error made by PWS is influenced by the slab thickness, in both
geometries studied, it turns out to be no more accurate in the case
of thin slabs than in the case of thick ones. This result confirms
that the fact that PWS does not take into account the screening of
electromagnetic fields cannot be the only explanation of the errors
made by this method. For silicon slabs, PWS was found to
underestimate or overestimate the Casimir interaction depending on
the thickness in both geometries, showing once again the complexity
of the error made by PWS.

\section{Acknowledgements}
The authors thank the ESF Research Networking Programme CASIMIR
(www.casimir-network.com) for providing excellent possibilities for
discussions and exchange.

\appendix

\section{Expressions for the successive primitive functions}

The successive primitive functions involved in the derivations of the different energies are:
\begin{align}
d(t) & =  -e^{-t} \left( \frac{1}{t} + \frac{4}{t^2} + \frac{20}{t^3} + \frac{48}{t^4} + \frac{48}{t^5} \right)
\end{align}
\begin{align}
e(t) & =  \Gamma(0,t) + 4 e^{-t} \left( \frac{1}{t^2} + \frac{3}{t^3} + \frac{3}{t^4} \right)
\end{align}
\begin{align}
f(t) & =  t \Gamma(0,t) - e^{-t} \left( 1+ \frac{4}{t^2} + \frac{4}{t^3} \right)
\end{align}
\begin{align}
g(t) & =  \left( \frac{t^2}{2} - 2 \right) \Gamma(0,t) +e^{-t} \left(-\frac{t}{2} + \frac{1}{2} + \frac{2}{t} + \frac{2}{t^2} \right)
\end{align}
\begin{align}
h(t) & =  \left( \frac{t^3}{6} -2t \right) \Gamma(0,t)+ e^{-t} \left(- \frac{t^2}{6} + \frac{t}{6} +\frac{5}{3} - \frac{2}{t} \right)
\end{align}
\begin{multline}
i(t)  =  \left( \frac{t^4}{24} -t^2+2 \right) \Gamma(0,t) \\
+ e^{-t} \left(- \frac{t^3}{24} + \frac{t^2}{24} +\frac{11t}{12} - \frac{3}{4} \right)
\end{multline}
\begin{multline}
j(t)  =  \left( \frac{t^5}{120} -\frac{t^3}{3}+2 t \right) \Gamma(0,t) \\
+ e^{-t} \left(- \frac{t^4}{120} + \frac{t^3}{120} +\frac{19t^2}{60} - \frac{17t}{60} - \frac{23}{15} \right) ~ .
\end{multline}
All these functions vanish in the $(t \rightarrow +\infty)$-limit, and diverge in the $(t \rightarrow 0)$-limit, except the last one, for which $j(0)=-\frac{23}{15}$.

\bibliographystyle{apsrev4-1}

%

\end{document}